# Auto-3DPFM: Automating Polarization-Vector Mapping at the Nanoscale

Ralph Bulanadi,[1,*] Marti Checa,[1,^] Michelle Wang,[2] Franck Rothen,[3] John Lasseter,[1] Sumner B. Harris,[1] Daniel Sando,[4,5] Valanoor Nagarajan,[4] Liam Collins,[1] Stephen Jesse,[1] Rama Vasudevan,[1] Yongtao Liu[1,#]

[1] Center for Nanophase Materials Sciences, Oak Ridge National Laboratory, Oak Ridge, TN 37831, USA

[2] Department of Electrical and Photonics Engineering, Technical University of Denmark, 2800 Kongens Lyngby, Denmark

[3] Department of Particle Physics, University of Geneva, Geneva 1205, Switzerland

[4] School of Materials Science and Engineering, UNSW Sydney, Kensington, New South Wales, 2052, Australia

[5] MacDiarmid Institute for Advanced Materials and Nanotechnology, School of Physical and Chemical Sciences, University of Canterbury, Christchurch 8410, New Zealand

* bulanadira@ornl.gov

^ checam@ornl.gov

# liuy3@ornl.gov

**Abstract**

The functional properties of ferroelectric materials are strongly influenced by ferroelectric polarization orientation; as such, access to consistent and precise characterization of polarization vectors is of substantial importance to ferroelectrics research. Here, we develop a fully automated three-dimensional piezoresponse force microscopy (Auto-3DPFM) technique automating all essential steps in interferometric PFM for 3D polarization vector characterization, including laser alignment, tip calibration and approach, image acquisition, polarization vector reconstruction, and visualization. The automation reduces the experimental burden of ferroelectric polarization vector characterization, while the back-and-forth calibration ensures consistency and reproducibility of 3D polarization reconstruction. An algorithmic workflow is also developed to identify domain walls and calculate their characteristic angles via a spatial vector-angle-difference method, presenting one unique capability enabled by Auto-3DPFM that is not accessible with traditional PFM techniques. Beyond representing a significant step forward in 3D polarization mapping, Auto-3DPFM promises to accelerate discovery via high-throughput and autonomous characterization in ferroelectric materials research. When integrated with machine learning and adaptive sampling strategies in self-driving labs, Auto-3DPFM will serve as a valuable tool for advancing ferroelectric physics and microelectronics development.

**Introduction**
The utility of ferroelectric materials is largely governed by the structure and dynamics of the underlying polarization order parameter. For most multi-axial ferroelectric materials, these polarizations arrange into complex domain patterns according to electrical and elastic boundary conditions, the material's grain structure, and a host of thermodynamic parameters including composition and temperature. Despite ferroelectrics being prevalent for over 100 years [1], structural imaging of ferroelectric domains at the nanoscale is a relatively recent innovation that has occurred in the past three decades [2]. Historically, domain structures have been imaged with larger-scale, polarized-light microscopy [3] that relied on anisotropic optical properties for contrast, with the accompanying resolution limits. Advances in microscopy methods have since made imaging the polarization of ferroelectric single crystals, ceramics, and thin films at nanoscale resolutions routine: scanning transmission electron microscopy (STEM) [4, 5] allows for direct imaging of atomic columns and therefore determination of the polarization vector through unit-cell-level structural analysis; while piezoresponse force microscopy (PFM) [6-8] can provide nanoscale-level (~10-1000 nm) imaging and spectroscopy of the material's polarization-dependent piezoelectric response. Unlike STEM, PFM has the added benefit of minimal sample preparation, and additionally, the ability to perform a wide range of spectroscopies that manipulate the domain structure [9-11], which although possible through in-situ STEM techniques [12], is considerably more challenging.

Although nanoscale domains have always been of interest (for example, consider the nanodomain mixtures of rhombohedral and tetragonal phases at the morphotropic phase boundary in lead zirconate titanate crystals [13]), the recent push towards understanding the rich tapestry of topological phases such as skyrmions, bubble domains, vortices and hedgehog states in ferroelectrics [14-16] has only intensified the need for accurate, high-veracity imaging of nanoscale domain patterns in ferroelectric materials. Manipulating and confirming the presence (or absence) of these states is critical to drive future nanoelectronics based on these novel architectures [17, 18]; however, the current method of confirming the vector orientation of the electric polarization vectors in PFM is both time-consuming and generally not quantitative. Vector PFM has traditionally required physically rotating the sample [19], which compounds errors when overlapping and comparing the sequentially-acquired data, and moreover, use of conventional optical beam deflection (OBD) makes quantitative studies extremely challenging. Recently, Proksch et al. developed a method to circumvent this problem by using an AFM with a quadrature-phase differential interferometer (QPDI) sensor and measuring PFM maps with the laser at four pre-defined locations, and were able to recover the full polarization vector [20].

In this work, we develop fully automated three-dimensional piezoresponse force microscopy (Auto-3DPFM), which produces consistent and reproducible three-dimensional mapping of ferroelectric polarization. This process integrates real-time algorithmic analysis of multimodal data with precise instrument control, automating all essential steps in interferometric PFM for nanoscale 3D polarization vector characterization, including laser alignment, tip calibration and approach, image acquisition, vector reconstruction and visualization. Beyond 3D polarization mapping, Auto-3DPFM also enables determination of domain-wall orientations that are inaccessible using traditional PFM techniques. This technique not only reduces experimental burden in 3D polarization characterization but also opens new opportunities for ferroelectric research, from nanoscale polarization characterization to autonomous materials discovery workflows, promising new steps forward in ferroelectric physics and next-generation microelectronic technologies.

**Results and Discussion**
*Vector PFM*
Traditional OBD-based vector PFM characterization relies on a series of measurements designed to track how the piezoresponse signal changes as the sample is rotated [19]. A shear deformation on the surface of the sample, $U$, can drive torsion of the cantilever which can be captured by the deflection of an optical beam (Fig. 1a). Rotating the sample allows for consequent rotation of this torsional axis, yielding sensitivity to different in-plane components, from which the true polarization vector can be reconstructed. Fig. 1b shows one such reconstruction on a thin film of bismuth ferrite, where polarization orientations visible in one orientation are not visible in the other (and vice versa); only through overlaying these components can the full orientation be determined. This rotation-based strategy is therefore essential for separating the various in-plane contributions that would otherwise remain mixed in a single measurement. In practice, the sample is physically rotated in defined increments, such as 45° or 90°. After each rotation, a new lateral PFM image is acquired. As the angle changes, both the amplitude and phase of the lateral response vary systematically, reflecting how the in-plane polarization projects onto the measurement axis. These angle-dependent lateral signals are then combined with the out-of-plane response to reconstruct the full 3D polarization vector. However, OBD-PFM is well known to suffer from significant measurement artifacts, as broadly discussed in previous works [21, 22], which can distort the polarization reconstruction and reduce accuracy. Moreover, the sample-rotation-based method itself poses practical challenges. It requires repeatedly locating the same measurement region after each rotation, typically within a sub-micrometer range; a process that is time-consuming and often impractical, especially when domain features are small, dense, or visually similar. These limitations greatly restrict the reliability and scalability of traditional vector OBD-PFM for high-throughput quantitative 3D polarization mapping.

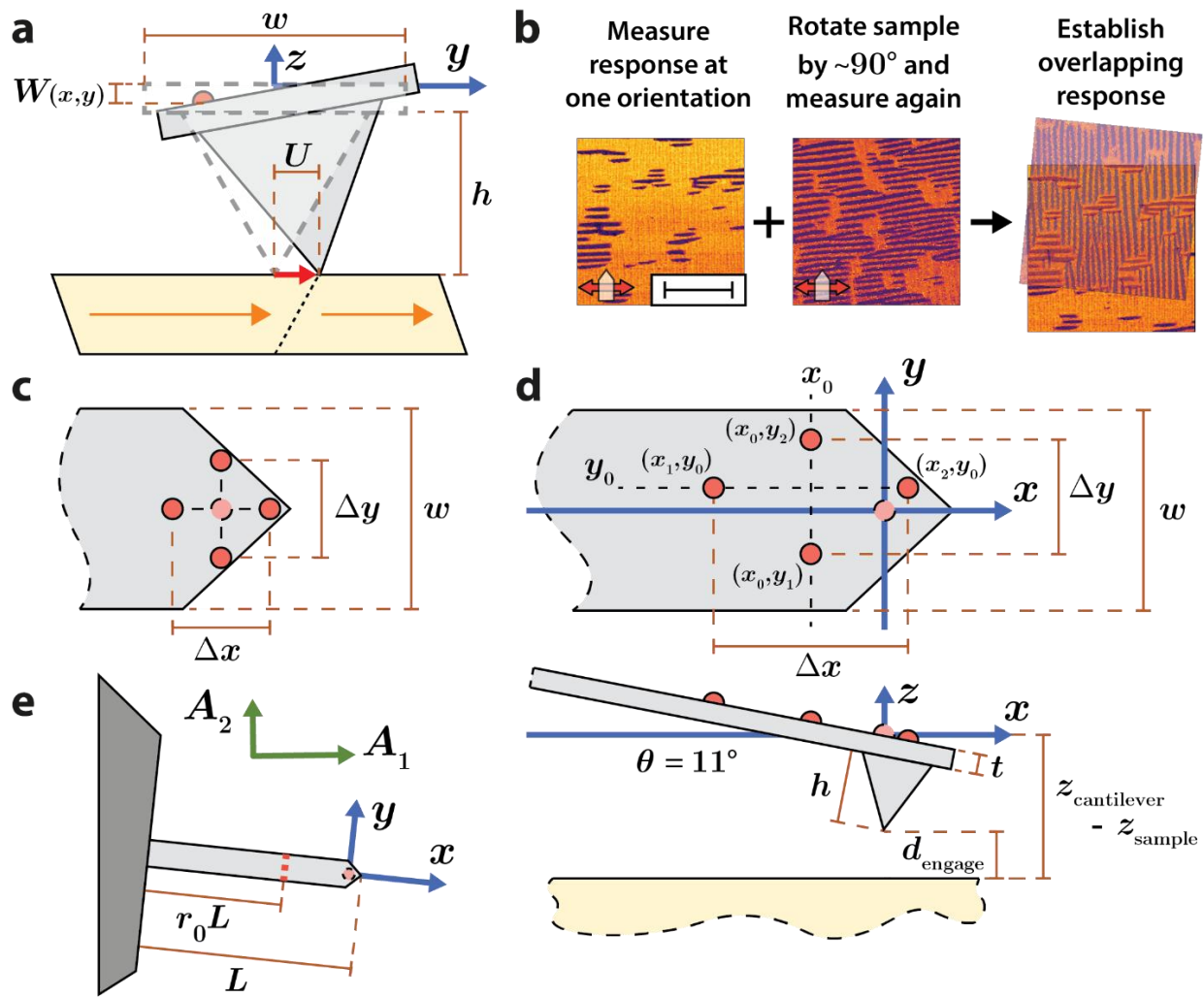

*Fig. 1. Acquiring vectoral polarization responses in PFM. a) Under the influence of an electric field by an AFM tip, the resulting surface vector **U** produces a position-dependent response in the cantilever. b) Traditional OBD-based systems require manual rotation that can yield full 3D reconstruction by establishing the overlapping response, here shown via phase-offset maps on a bismuth ferrite thin film. c) As described in Ref. [20], the in-plane polarization response can be determined by positioning an interferometric laser symmetrically about the blind spot. d) [Upper] Top and [Lower] side views of this present approach, allowing for larger values of $\Delta x$, $\Delta y$. e) A wider view, showing the relationship between $(x, y)$ cantilever axes and $(A_1, A_2)$ piezoelectric-actuator axes. Scalebar: (b) 2 µm.*

Recent advances in interferometric PFM have shown that it is possible to recover all three orthogonal piezoresponse components using only four images [20]. Interferometric PFM directly measures tip displacement rather than cantilever slope and (if used correctly) minimizes electrostatic and mechanical artifacts compared to OBD-PFM. This reduces ambiguity in interpreting signals and enables more reliable quantitative comparisons across measurements.

In addition, interferometric vector PFM avoids manual sample rotation, removing the need to repeatedly realign the sample to the same measurement region—one of the major challenges in rotation-based OBD-PFM vector mapping. This advantage arises from assuming the cantilever and tip act as a rigid body; as shown in Fig. 1a, the surface vector $U$ leads to a position-dependent vertical displacement $W(x,y) = w(x,y)e^{i\phi_w(x,y)}$, where $\phi$ denotes the phase of the response. In earlier work [20], the effect of this response on the cantilever is described as:

$$W(x,y) \approx \frac{x}{h}U_x + \frac{y}{h}U_y + \left(1 + \frac{x}{r_0 L}\right)U_z \qquad 1$$

where $U_j = u_j e^{i\phi_j}$ for $j = x, y, z$ is the motion at the tip–sample contact (typically underneath the "blind spot" at $(0,0)$), $r_0 \approx 2/3$ is the dimensionless location of the effective lever arm for the zeroth mode, $L$ is the cantilever length, and $h$ is the height of the tip. These parameters are shown diagrammatically through Figs. 1a, d, and e. By measuring the cantilever motion $W(x,y)$ at a minimum of three non-collinear distinct positions $(x,y)$, one can fully reconstruct $U$. In earlier work [20] this is achieved using four measurement points located symmetrically with respect to the tip blind spot at $(\pm x, 0)$ and $(0, \pm y)$ (Fig. 1c).

In practice, however, positioning the laser precisely relative to the blind spot is experimentally challenging and limits the potential range of measurement positions, particularly when the blind spot is close to the edge. Moreover, because the laser-positioning actuators are not perfectly linear or repeatable, automatically driving the laser to the exact desired spot can be difficult and often requires a slow, iterative alignment process. To overcome these limitations, we express the problem in terms of relative separations between measurement points, which are easier to control experimentally (Fig. 1d). Let $\Delta x = x_2 - x_1$ and $\Delta y = y_2 - y_1$, denote respectively the horizontal and vertical measurement separations, where the absolute positions $x_{1,2}$ and $y_{1,2}$ relative to the blind spot at $(0,0)$ remain uncertain. For notational convenience, we also express the cantilever motion in terms of these separations, defining $\Delta W_x := W(x_2, y_0) - W(x_1, y_0)$ and $\Delta W_y := W(x_0, y_2) - W(x_0, y_1)$. We then obtain the following relations from Eq. 2:

$$\begin{aligned} U_x &\approx \frac{h}{\Delta x}\Delta W_x - \frac{h}{r_0 L}U_z \\ U_y &\approx \frac{h}{\Delta y}\Delta W_y \\ U_z &\approx W(x,y) - \frac{x}{\Delta x}\Delta W_x - \frac{y}{\Delta y}\Delta W_y \end{aligned} \qquad 2$$

Assuming the tip height $h$ and cantilever length $L$ are known, $U_y$ is determined uniquely while the remaining components, $U_x$ and $U_z$, require the coordinates of one measurement with respect to the blind spot. To manage this, we compute $U_x$ and $U_z$ by averaging over the four measurement points to reduce the sensitivity to positioning uncertainty and provide a more reliable estimate of the motion beneath the tip.

*Automating Three-Dimensional PFM*

Building on the mechanism described above, we establish a workflow that enables fully automated and consistent 3D PFM characterization (Fig. 2). The process starts by loading the sample, aligning the interferometric laser with the cantilever tip, and preparing the probe to engage onto the surface (Initial Setup in Fig. 2), which are the essential steps in virtually all microscopy experiments. At this stage, basic cantilever properties such as its nominal width are recorded.

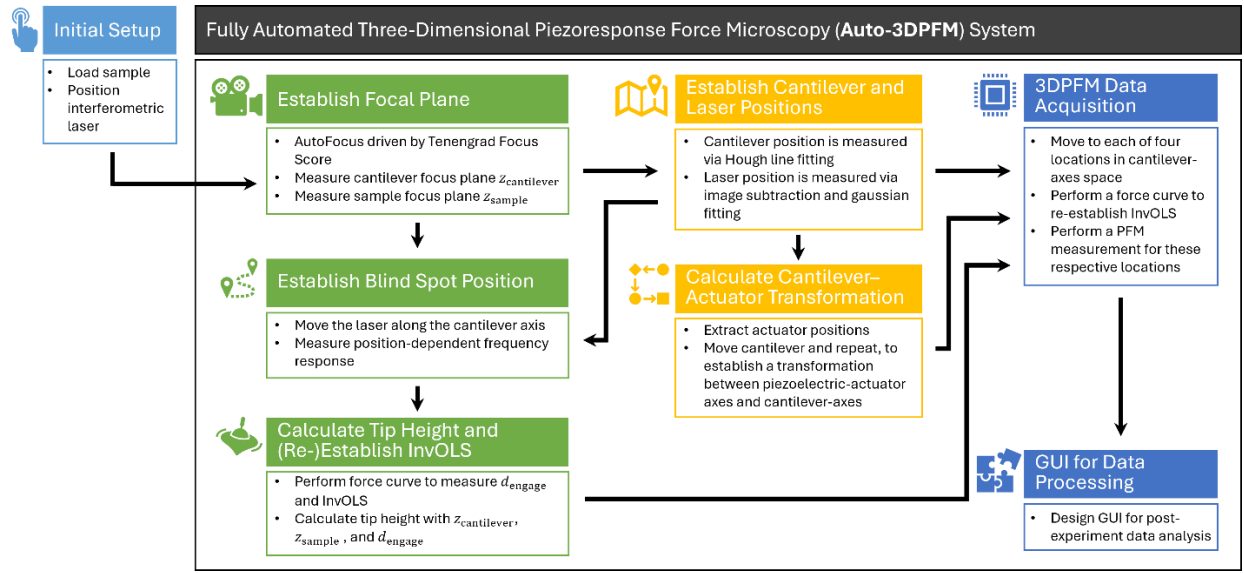

**Fig. 2.** *The workflow of Auto-3DPFM. The automation reduces the experimental burden in 3D polarization mapping and the back-and-forth calibration ensures the accuracy and reproducibility of 3D polarization reconstruction. A recording of this process is shown in Supplementary Video 1.*

After the initial setup, all steps for 3D polarization characterization mapping are automated (Fig. 2). The system first automatically determines the focal planes of both the cantilever and the sample using an AutoFocus driven by a Tenengrad focus score (Fig. 3a) [23, 24], allowing establishment of their relative vertical positions. The workflow identifies the cantilever orientation and measures the laser spot position, which together define the sample's location within the imaging coordinates. The system also records the response of the internal piezoelectric actuators. By repeating these steps while moving the cantilever, the workflow constructs a mapping between the actuator coordinate system and the physical cantilever axes (denoted as $(A_1, A_2)$ and $(x, y)$ respectively in Fig. 1e). The blind-spot position is then located by scanning the laser along the cantilever axis and monitoring the corresponding tip response (Fig. 3b). Once this reference point is identified, the system can position the laser with high precision. A force curve is taken at the blind spot to determine the engagement distance and extract key parameters such as the tip height and inverse optical lever sensitivity (InvOLS) (Fig. 3c).

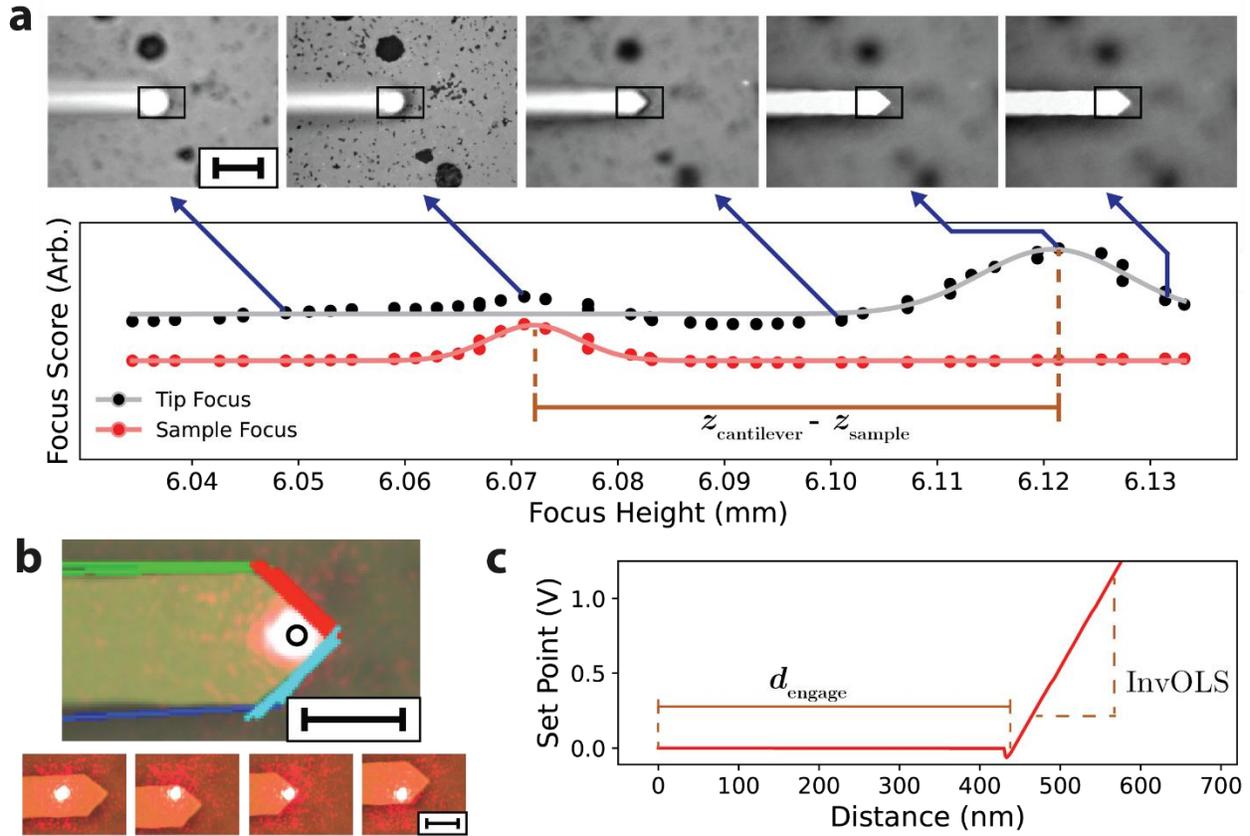

***Fig. 3.*** *Automated processes used in the Auto-3DPFM. a) AutoFocus using a Tenengrad focus score. [Upper] shows a few representative sections of the optical capture, where the inner rectangle describes the tip-focus region. [Lower] plots the Tenengrad focus score as a function of focus height. b) [Upper] computer vision of the cantilever, showing the cantilever outline and laser diode position. [Lower] shows controlled laser positions at distinct locations. c) Force–Distance curve, showing how $d_{\text{engage}}$ and InvOLS can be calculated. Scalebar: (a) 50 μm; (b) 20 μm.*

With all reference parameters in place, the workflow sequentially moves the laser to four designated positions relative to the cantilever. At each location, the system updates InvOLS through a force measurement and then acquires a PFM image with the laser at that pre-set position. If necessary, the topography image can be used to correct for small drift to ensure that all images refer to the same exact physical region of the sample [25]. Finally, once all measurements are collected, the workflow reconstructs the full vector response using Eq. 2. The resulting 3D polarization components can then be stored for further analysis and visualization; we here use the Hierarchical Data Format 5 (HDF5) file format, which has a history of being used for the storage of such multidimensional scanning probe microscopy datasets [26, 27]. For additional details on how each step is enabled in the automated workflow, including machine learning methods for analysis and programmatic instrument control, see Supplementary Information. A recording of the full process is shown in Supplementary Video 1, while a recording of the AutoFocus is shown in Supplementary Video 2.

While Auto-3DPFM has been designed to operate without human intervention and immediately provide access to polarization-vector mapping, certain aspects of data analysis, specific research objectives, and the discovery of new ferroelectric physics could benefit from post-experiment analysis. To facilitate post-experiment analysis, user-friendly and interactive graphical user interfaces (GUIs) have also been developed, as shown in Fig. 4. Fig. 4a shows a GUI facilitating manual overlay correction with controllable color and opacity for each image to ensure accuracy in the final polarization-vector maps. Fig. 4b shows a GUI that calculates the vector-angle difference between specified points. Once a point of interest (reference point) is specified by the user, the colormap plots the vector-angle difference between all points against the point of interest. The GUI also provides the numerical value of a second point (specified by users) against the reference point. Additional scaling factors can optionally be used to allow for potential changes between torsional and out-of-plane tip mechanics.

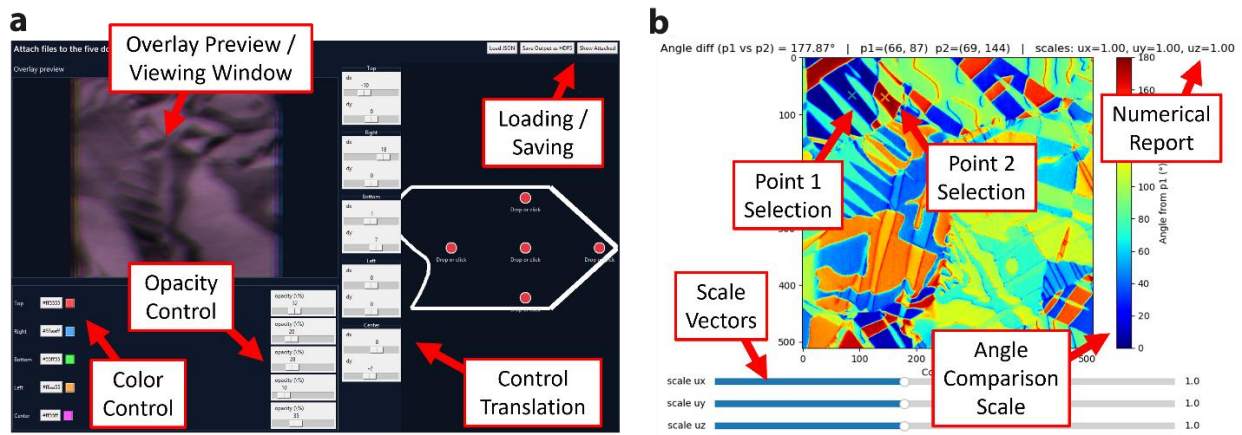

*Fig. 4. Interactive GUIs to facilitate post-experiment analysis, supporting specific research objectives and novel discoveries. a) Overlay correction. b) Vector-Angle difference calculation.*

*Auto-3DPFM Results*
Example results from Auto-3DPFM on a sample of polycrystalline, rhombohedral lead zirconate titanate (PZT) are shown in Fig. 5 (Figs. S1, S2 provide additional characterization of this sample). The domains of such a polycrystalline sample allow for diversity in measured polarization directions, while still being bounded by predictable domain-wall configurations and grain boundaries, and so presents a suitable model sample for polarization mapping.

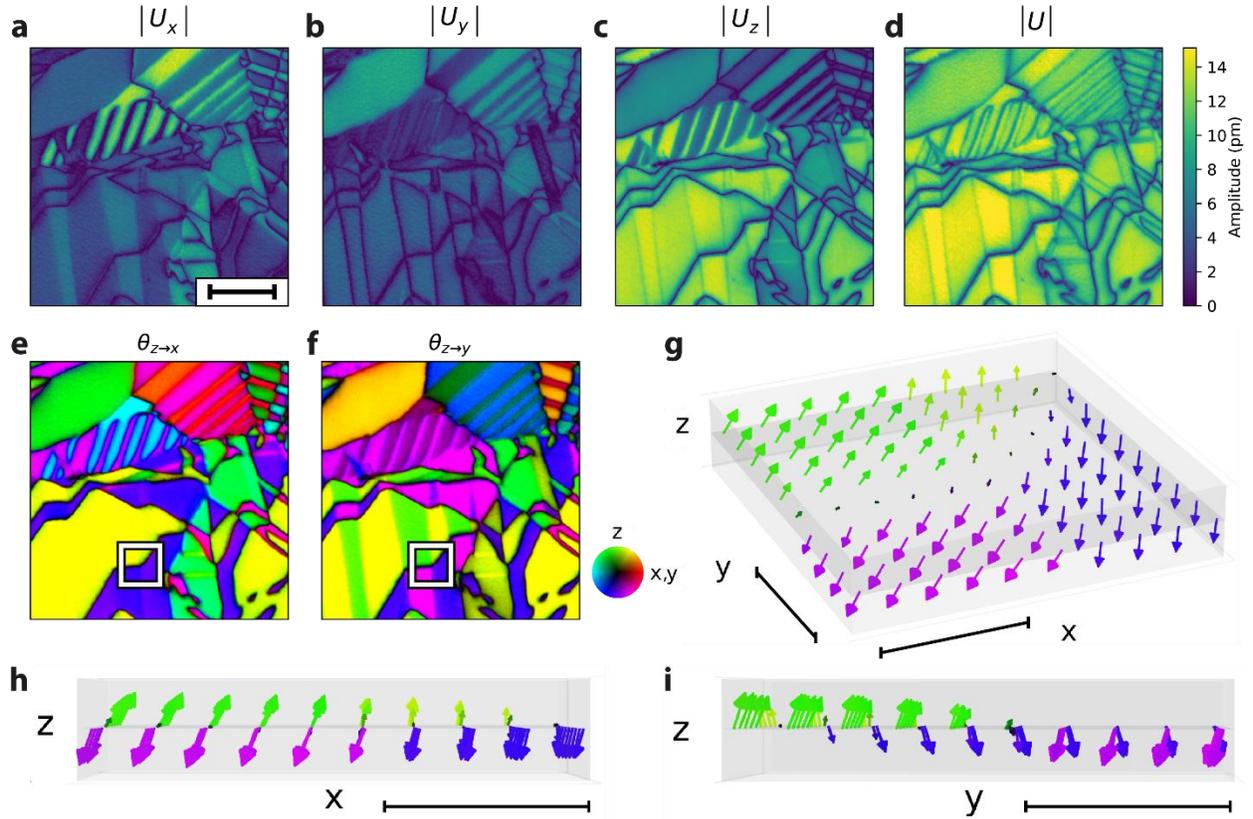

***Fig. 5.*** *Auto-3DPFM on rhombohedral PZT. Magnitude of the surface displacement vector components are shown in a) x, b) y, and c) z axes. Overall magnitude is shown in d). The polarization angles are shown along the e) x–z and f) y–z planes. A zoomed-in vector field plot of the area in the white squares is shown in g), with the lateral view along the x–z (h) and y–z (i) planes. Scalebar: (a) 2 μm; (g–i) 500 nm.*

In Fig. 5(a–c), the amplitude of the response in the three orthogonal components $|U_x|$, $|U_y|$, and $|U_z|$, is shown. We observe domains dominated by an out-of-plane response, especially towards the lower-left of the image, bounded by associated domain walls. However, towards the upper-right of the image, generally reduced out-of-plane responses are observed. By supplementing the out-of-plane component $U_z$ with measurements of the in-plane components $U_x$ and $U_y$, more complete characterization can be achieved. In the upper-right region, the reduced $|U_z|$ corresponds with an increased $|U_x|$ and $|U_y|$, as such a plot of $|U|$ in Fig. 5d reveals a generally uniform vector magnitude observed in each domain of the sample. A map of the angle of $U$ along the x–z and y–z planes are shown in Figs. 5(e, f), with a small section displayed as a vector field in Figs. 5(g–i). Even in this section, a diversity of domain and domain-wall configurations are present but would otherwise be indistinguishable in a map of $U_z$ alone.

This rapid, reproducible method to execute and analyze such a measurement would be of immense use to future research in ferroelectric materials. The remainder of this paper showcases examples of practical scientific use cases enabled by such high-throughput, automated analysis.

*Auto-3DPFM Force-Dependent and Cantilever-Angle-Dependent Studies*

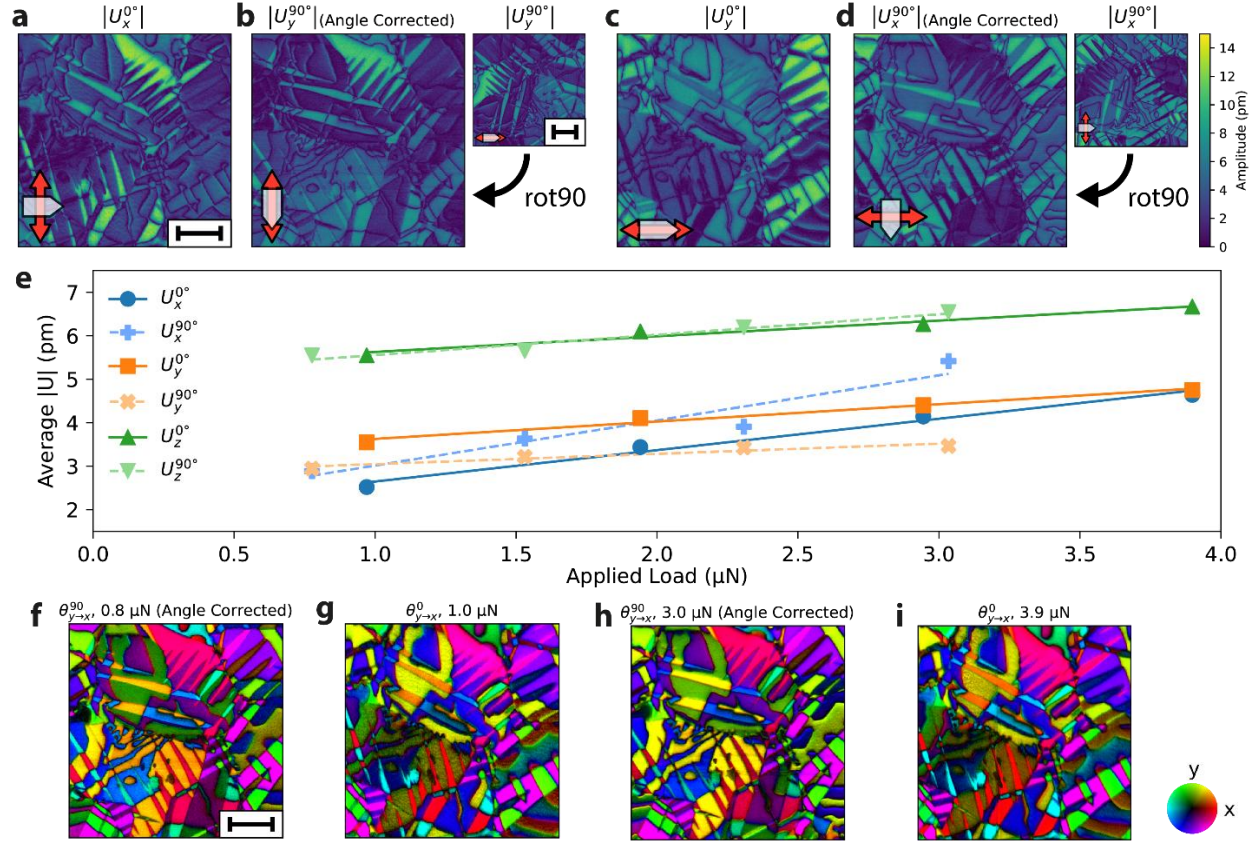

**Fig. 6.** Force- and cantilever-angle-dependent effects on Auto-3DPFM. Maps of the response a) in the direction of the cantilever, b) perpendicular to the cantilever after sample rotation by 90°, showing similar data. c) shows the response perpendicular to the cantilever, while d) shows response along the cantilever axis after sample rotation by 90°, again showing similar data. e) shows the average piezoresponse magnitude at different forces and rotations. f–i) show the polarization angle in the x–y plane at different forces and rotations, showing qualitatively similar mappings. Scalebar: 2 μm.

Ferroelectric polarization is strongly influenced by stress and strain [28, 29] which can promote the formation and enhancement of a specific ferroelectric phase, especially in materials with compositions near a morphotropic phase boundary. Flexoelectric effects can also inhibit or enhance a particular ferroelectric direction, leading to complex tribological phenomena and emergent physics [30, 31]. For PFM in particular, torsion and slipping of the tip could lead to force-dependent artefacts under low-stress boundary conditions at the tip–sample junction. In providing a fully automated method to measure polarization under set stresses, while also updating the system InvOLS before each scan to ensure a consistent applied force, Auto-3DPFM provides a unique method to investigate the influence of stresses on ferroelectric polarization vectors. Auto-3DPFM was performed with an applied load ranging from 0.8 µN to 4 µN, at two different orientations spaced approximately 90° apart. The magnitude of the in-plane responses for the first orientation, denoted as $|U_x^{0°}|$ and $|U_y^{0°}|$, are shown in Figs. 6(a, c). The responses from the second orientation, $|U_x^{90°}|$ and $|U_y^{90°}|$, are shown in Figs. 6(d, b), respectively. Note once each of these images are rotated by 90° to reverse the physical sample rotation, the map of $|U_x^{0°}|$ is highly comparable to the measurement of $|U_y^{90°}|$, and likewise the measurement of $|U_y^{0°}|$ is similar to a measurement of $|U_x^{90°}|$, proving the reproducibility of this Auto-3DPFM technique.

To measure the force-dependence of the measured response, an average of the measured oscillation amplitude along each axis was taken over each image and compared (Fig. 6e). An increase in applied load slightly increased the measured oscillation amplitude for each vector component: after linear fitting ($R^2 > 0.9$ for all lines) both $|U_y|$ and $|U_z|$ increase by 15 % and 17% respectively with an increase in applied load from 1 µN to 3 µN. We attribute this to improved tip–sample contact. However, the oscillation amplitude in $|U_x|$ (in the direction of the cantilever axis) increased by 69% along the same force range, suggesting either sample-dependent effects, or an artefact related to behavior along the cantilever axis. A similar trend was observed in the measurements taken when the sample was rotated. Here, $|U_y^{90°}|$ and $|U_z^{90°}|$ increased by 22% and 13% respectively when the applied load increased from 1 µN to 3 µN, consistent with the changes observed in of $|U_y^{0°}|$ and $|U_z^{0°}|$; however, $|U_x^{90°}|$ increased by a dramatically higher 55%, akin to the increased observed in $|U_x^{0°}|$. This suggests a uniquely force-dependent artefact related to measured components along the cantilever axis.

However, the vectors $U_x^{0°}$ and $U_y^{90°}$ refer to the same polarization vector on the sample itself; likewise, $U_x^{90°}$ and $U_y^{0°}$ should also be equal. A crossover between these respective pairs of lines occur at 1.96 µN and 1.82 µN. This suggests that a measurement at ~1.9 µN would be self-consistent, and act as a "force blind spot" to nullify the cantilever-axis artefacts observed. It is also notable that, at this "force blind spot", the measured piezoresponse in the out-of-plane direction remains substantially higher than either of the in-plane piezoresponses, and this was generally observed throughout the sample. While the sample is polycrystalline and therefore would be expected to have an overall isotropic polarization response, an out-of-plane response may be preferred energetically due to either mechanical effects from polishing [32], or from screening charges and other associated surface electrostatics [33]. Alternatively, this may also suggest additional artefacts, where torsional distortions on the cantilever are inhibited by a higher spring constant than an out-of-plane deflection [34]. Despite these potential artefacts, regardless of force-dependence or cantilever-orientation, the measured domain polarizations remain generally consistent (Fig. 6f–i; quantitative comparison between these maps are shown in Fig. S3).

Additional studies to determine potential sample-dependent and tip-dependent effects on this "force blind spot", or apparent out-of-plane preference, could lead to further numerical precision.

*Auto-3DPFM Domain Wall Analysis*
Characterization of various domain orientations allows the corresponding domain-wall orientations to be identified. This identification is of paramount importance for further high-throughput autonomous studies of domain walls. Here, we further develop a method to identify domain-wall positions and measure their orientations from the Auto-3DPFM output (Fig. 7a). This method can be integrated into autonomous experiment systems for accelerating scientific discovery [35], where the experiment can be expressly driven to investigate unique or novel domain and domain-wall structures [36].

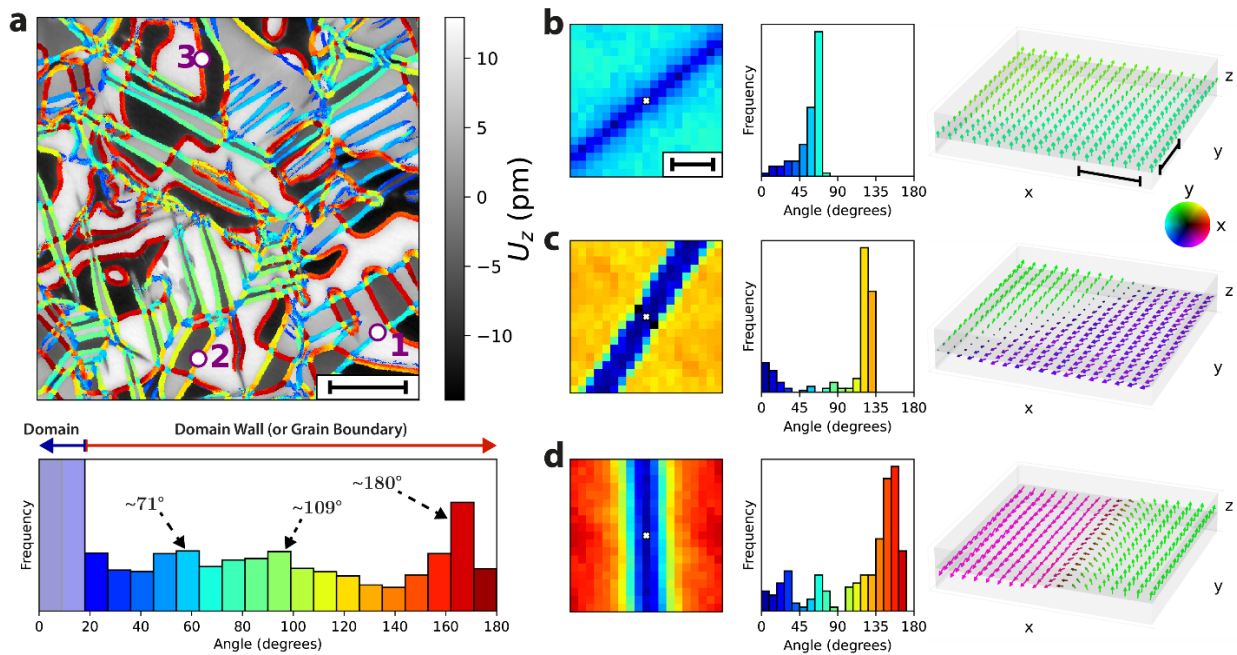

*Fig. 7.* Extraction of domain-wall orientation from 3D-PFM. a) [Upper] Final domain-wall orientation overlaid onto the $U_z$ response; color bar for domain-wall orientation is shown in the [Lower] histogram. b–d) show the domain-wall analysis for a ~71°, ~109°, and ~180° domain wall respectively. [Left] shows a map of vector-angle differences in the vicinity of the analyzed point, [Middle] shows a histogram of distributions in this vicinity, and [Right] shows a quiver plot of the polarization vectors. Scalebar: (a) 2 μm; (b–d) 100 nm.

To identify these domain-wall orientations, the vector-angle differences in the neighborhood of each point are measured and compared. In detail, for every pixel $(i,j)$, we examine all pairs of points located symmetrically about it at $(i + \Delta i, j + \Delta j)$ and $(i - \Delta i, j - \Delta j)$, where $\Delta i$ and $\Delta j$ independently take integer values in the range [-9, 9]. For each such pair, the difference between their vector angles is computed. The domain-wall angle assigned to the point $(i,j)$ is then the most frequent vector-angle difference in this neighborhood, determined using 9° histogram bins. Regions with a most-frequent vector-angle difference >18° are classified as domain walls; regions that have a lower characteristic vector-angle difference are classified as domains. Notably, previous domain-wall detection methods rely on supervised machine-learning approaches that require substantial effort in data annotation for model training [37]. In contrast, Auto-3DPFM domain-wall analysis eliminates the need for annotation and is readily available in the Auto-3DPFM package; this method could further be extended for polarization vector variation analysis across grain boundaries in the future [38].

According to the analysis, domain-wall orientations at ~71°, ~109°, and ~180° have a relatively higher occurrence, as expected for a material with a rhombohedral crystallographic structure. A map showing these vector-angle differences are shown on the left sides of Figs. 7b–d and marked as points (1), (2) and (3) respectively on Fig. 7a. Histograms showing the angle occurrence around these points are shown in the middle of these subfigures; quiver-plot representations of these domain walls are shown in the right sides. Notably, the peaks attributed to 71°, 109° and 180° domain walls in Fig. 7a tend to be lower than their nominal values; this is attributed to this calculation including points that lie on the domain wall, and on the same side of the domain wall. These are observed as dark-blue bands seen on the left sides of Figs. 7b–d. This could be increased by increasing the maximum values of $\Delta i$, $\Delta j$, but this would reduce calculation speed and would reduce the capability to measure walls around smaller domains.

**Conclusions**
In summary, we developed an Auto-3DPFM technique that enables automated characterization of three-dimensional polarization vectors, significantly reducing the experimental burden associated with mapping complex ferroelectric polarization vectors. This technique represents the current state-of-the-art in polarization vector mapping, demonstrating low noise, high reproducibility, and substantial information gain in ferroelectric vector characterization. The measured polarization orientations are consistent and robust across various experimental conditions such as different applied forces and sample rotations. Comprehensive polarization vector information from Auto-3DPFM can be used to determine domain-wall orientations, which are not accessible using traditional PFM techniques. This capability opens a broad range of possibilities for future ferroelectric research.

Auto-3DPFM particularly enables new opportunities for self-driving labs in ferroelectric research. The ability to reliably and automatically map 3D polarization aligns with broader efforts toward autonomous experimentation and accelerated discovery. When integrated with machine-learning-driven experimental planning, adaptive sampling strategies, and closed-loop optimization approaches, Auto-3DPFM can enable intelligent interrogation of ferroelectric materials. This capability is particularly valuable for precisely characterizing specific features (e.g., domains, domain walls, and grain boundaries), understanding switching dynamics, and discovering composition–structure–property relationships in complex ferroelectric systems.

Furthermore, Auto-3DPFM can be implemented on typical commercial atomic force microscopes (AFM) equipped with a QPDI detector. This compatibility with standard commercial AFM platforms ensures broad accessibility, facilitating its adoption across the ferroelectrics research community and enabling wide integration into emerging autonomous materials discovery platforms.

**Acknowledgements:**
This research and workflow development was sponsored by the INTERSECT Initiative as part of the Laboratory Directed Research and Development Program of Oak Ridge National Laboratory, managed by UT-Battelle, LLC for the US Department of Energy under contract DE-AC05-00OR22725. M.C acknowledges the support from Laboratory Directed Research and Development Program of Oak Ridge National Laboratory, managed by UT-Battelle, LLC, for the US Department of Energy. Piezoresponse force microscopy was performed at and supported by the Center for Nanophase Materials Sciences (CNMS), which is a US Department of Energy, Office of Science User Facility at Oak Ridge National Laboratory.

**Conflicts of Interest:**
The authors declare no conflict of interest.

**Data and Code Availability**
The data and code presented in this work is available on GitHub [39].

**References**

1. Whatmore, R.W., et al., *100 years of ferroelectricity—A celebration.* APL Materials, 2021. **9**(7).
2. Gruverman, A., O. Auciello, and Tokumoto, *Scanning force microscopy for the study of domain structure in ferroelectric thin films.* Journal of Vacuum Science & Technology B: Microelectronics and Nanometer Structures Processing, Measurement, and Phenomena, 1996. **14**(2): p. 602-605.
3. Schmid, H., *Polarized light microscopy (PLM) of ferroelectric and ferroelastic domains in transmitted and reflected light*, in *Ferroelectric Ceramics: Tutorial reviews, theory, processing, and applications.* 1993, Springer. p. 107-126.
4. Nelson, C.T., et al., *Domain dynamics during ferroelectric switching.* Science, 2011. **334**(6058): p. 968-971.
5. Borisevich, A., et al., *Quantitative aberration-corrected STEM for studies of oxide superlattices and topological defects in layered ferroelectrics.* Microscopy and Microanalysis, 2020. **26**(S2): p. 1194-1195.

6. Eng, L., et al., *Ferroelectric domain characterisation and manipulation: a challenge for scanning probe microscopy.* Ferroelectrics, 1999. **222**(1): p. 153-162.
7. Gruverman, A. and S.V. Kalinin, *Piezoresponse force microscopy and recent advances in nanoscale studies of ferroelectrics.* Journal of materials science, 2006. **41**(1): p. 107-116.
8. Kalinin, S.V., A. Rar, and S. Jesse, *A decade of piezoresponse force microscopy: progress, challenges, and opportunities.* IEEE transactions on ultrasonics, ferroelectrics, and frequency control, 2006. **53**(12): p. 2226-2252.
9. Balke, N., et al., *Exploring local electrostatic effects with scanning probe microscopy: implications for piezoresponse force microscopy and triboelectricity.* ACS nano, 2014. **8**(10): p. 10229-10236.
10. Jesse, S., A.P. Baddorf, and S.V. Kalinin, *Switching spectroscopy piezoresponse force microscopy of ferroelectric materials.* Applied physics letters, 2006. **88**(6).
11. Jesse, S., et al., *The band excitation method in scanning probe microscopy for rapid mapping of energydissipation on the nanoscale.* Nanotechnology, 2007. **18**(43): p. 435503.
12. Cai, S., et al., *In situ observation of domain wall lateral creeping in a ferroelectric capacitor.* Advanced Functional Materials, 2023. **33**(50): p. 2304606.
13. Rossetti, G., et al., *Ferroelectric solid solutions with morphotropic boundaries: Vanishing polarization anisotropy, adaptive, polar glass, and two-phase states.* Journal of Applied Physics, 2008. **103**(11).
14. Yadav, A., et al., *Observation of polar vortices in oxide superlattices.* Nature, 2016. **530**(7589): p. 198-201.
15. Das, S., et al., *Observation of room-temperature polar skyrmions.* Nature, 2019. **568**(7752): p. 368-372.
16. Govinden, V., et al., *Spherical ferroelectric solitons.* Nature Materials, 2023. **22**(5): p. 553-561.
17. Checa, M., et al., *On-demand nanoengineering of in-plane ferroelectric topologies.* Nature Nanotechnology, 2025. **20**(1): p. 43-50.
18. Checa, M., et al., *Autonomous Multistate Nanoencoding Using Combinatorial Ferroelectric Closure Domains in BiFeO3.* ACS nano, 2025. **19**(30): p. 27692-27701.
19. Kalinin, S.V., et al., *Vector piezoresponse force microscopy.* Microscopy and Microanalysis, 2006. **12**(3): p. 206-220.
20. Proksch, R. and R. Wagner, *3D Vector Piezoresponse Imaging with Interferometric Atomic Force Microscopy.* Small Methods, 2025: p. 2401918.
21. Collins, L., et al., *Quantitative electromechanical atomic force microscopy.* ACS nano, 2019. **13**(7): p. 8055-8066.
22. Vasudevan, R.K., et al., *Ferroelectric or non-ferroelectric: Why so many materials exhibit "ferroelectricity" on the nanoscale.* Applied Physics Reviews, 2017. **4**(2).
23. Krotkov, E., *Focusing.* International Journal of Computer Vision, 1988. **1**(3): p. 223-237.
24. Pech-Pacheco, J.L., et al. *Diatom autofocusing in brightfield microscopy: a comparative study.* in *Proceedings 15th International Conference on Pattern Recognition. ICPR-2000.* 2000. IEEE.
25. Gaponenko, I., et al., *Computer vision distortion correction of scanning probe microscopy images.* Scientific reports, 2017. **7**(1): p. 669.
26. Musy, L., et al., *Hystorian: A processing tool for scanning probe microscopy and other n-dimensional datasets.* Ultramicroscopy, 2021. **228**: p. 113345.

27. Vasudevan, R.K., et al., *A processing and analytics system for microscopy data workflows: the pycroscopy ecosystem of packages.* Advanced Theory and Simulations, 2023. **6**(11): p. 2300247.
28. Liu, Y., et al., *Stress and curvature effects in layered 2D ferroelectric CuInP2S6.* ACS nano, 2023. **17**(21): p. 22004-22014.
29. Morozovska, A.N., et al., *The strain-induced transitions of the piezoelectric, pyroelectric, and electrocaloric properties of the CuInP2S6 films.* AIP Advances, 2023. **13**(12).
30. Checa, M., et al., *Nanoscale Polarization-Dependent Young's Modulus of Ferroelectric BaTiO3 Single Crystals.* ACS nano, 2025. **19**(10): p. 9835-9843.
31. Cho, S., et al., *Switchable tribology of ferroelectrics.* Nature communications, 2024. **15**(1): p. 387.
32. Cheng, S., I.K. Lloyd, and M. Kahn, *Modification of surface texture by grinding and polishing lead zirconate titanate ceramics.* Journal of the American Ceramic Society, 1992. **75**(8): p. 2293-2296.
33. Genenko, Y.A., O. Hirsch, and P. Erhart, *Surface potential at a ferroelectric grain due to asymmetric screening of depolarization fields.* Journal of Applied Physics, 2014. **115**(10).
34. Pettersson, T., et al., *Comparison of different methods to calibrate torsional spring constant and photodetector for atomic force microscopy friction measurements in air and liquid.* Review of Scientific Instruments, 2007. **78**(9).
35. Liu, Y., et al., *AEcroscopy: a software–hardware framework empowering microscopy toward automated and autonomous experimentation.* Small Methods, 2024. **8**(10): p. 2301740.
36. Bulanadi, R., et al., *Beyond optimization: Exploring novelty discovery in autonomous experiments.* ACS Nanoscience Au, 2025.
37. Liu, Y., et al., *Exploring physics of ferroelectric domain walls in real time: deep learning enabled scanning probe microscopy.* Advanced Science, 2022. **9**(31): p. 2203957.
38. Liu, Y., et al., *Disentangling electronic transport and hysteresis at individual grain boundaries in hybrid perovskites via automated scanning probe microscopy.* ACS nano, 2023. **17**(10): p. 9647-9657.
39. Bulanadi, R., *Auto-3DPFM.* 2025: GitHub.